\documentclass[preprint2]{aastex}
\def\stacksymbols #1#2#3#4{\def\theguybelow{#2}
	\def\verticalposition{\lower#3pt}
	\def\spacingwithinsymbol{\baselineskip0pt\lineskip#4pt}
	\mathrel{\mathpalette\intermediary#1}}
\def\intermediary #1#2{\verticalposition\vbox{\spacingwithinsymbol
	\everycr={}\tabskip0pt
	\halign{$\mathsurround0pt#1\hfil##\hfil$\crcr#2\crcr
		\theguybelow\crcr}}}
\def\lta{\stacksymbols{<}{\sim}{2.5}{.2}}
\def\gta{\stacksymbols{>}{\sim}{3}{.5}}

\begin{document}
\title{COOLING FLOW STAR FORMATION AND THE APPARENT STELLAR AGES 
OF ELLIPTICAL GALAXIES}

\author{William G. Mathews$^2$ and Fabrizio Brighenti$^{2,3}$}

\affil{$^2$University of California Observatories/Lick Observatory,
Board of Studies in Astronomy and Astrophysics,
University of California, Santa Cruz, CA 95064\\
mathews@lick.ucsc.edu}

\affil{$^3$Dipartimento di Astronomia,
Universit\`a di Bologna,
via Ranzani 1,
Bologna 40127, Italy\\
brighenti@bo.astro.it}






\clearpage

\vskip .2in

\begin{abstract}

Simple theoretical arguments indicate that cooled interstellar gas 
in bright 
elliptical galaxies forms into a young stellar population having 
a bottom-heavy, but optically luminous IMF extending to 
$\sim 2$ $M_{\odot}$. When the colors and spectral features of this
young population are combined with those
of the underlying old stellar population, 
the apparent ages are significantly reduced,
similar to the relatively young apparent ages observed in many ellipticals. 
Galactic mergers are not required to resupply young stars. 
The sensitivity of 
continuous star formation to $L_B$ and $L_x/L_B$ is likely to account 
for the observed spread in apparent ages among elliptical galaxies. 
Local star formation is accompanied by enhanced 
stellar H$\beta$ equivalent widths, 
stronger optical emission lines, 
more thermal X-ray emission and lower 
apparent temperatures in the hot gas. The young stars should cause 
$M/L$ to vary with galactic radius, perturbing the fundamental plane of the
old stars alone.

\end{abstract}


\clearpage

\section{INTRODUCTION}

Traditionally, elliptical galaxies have been regarded as ancient stellar
systems in which evolutionary processes have been exhausted or completely
arrested. However, recent observational and theoretical developments have led
to a reassessment of both the ages of the stars in ellipticals and the 
ages since the stellar system merged into its current configuration.
Our objective here is to illustrate the possibly important
contribution to the apparent global stellar age in ellipticals 
due to a population 
of young, intermediate mass stars formed from cooled interstellar 
(cooling flow) gas. We also review the related structural history
of ellipticals.
The high degree of structural regularity among ellipticals has been 
used to argue that they were among the first galaxies that formed.
Of most interest are the remarkable thinness of the fundamental plane
(Djorgovski \& Davies 1987; Dressler et al. 1987; Renzini \& Ciotti 1993),
the tightness of the color magnitude relation (Bower, Lucey \& Ellis 1992)
and the tight correlation between central $M_{g2}$ line strength and 
central velocity dispersion $\sigma$ (Bender, Burstein, \& Faber 1993;
Ziegler \& Bender 1997), indicating an intimate connection between 
parameters that characterize chemical and dynamical evolution. 
The small scatter about these relations is difficult to reproduce 
if there is a wide range of ages among ellipticals. 
Moreover, distant elliptical galaxies at redshifts $z \gta 1$ exhibit
these same strong correlations (Aragon-Salamanca et al. 1993;
Ellis et al. 1997; Stanford, Eisenhardt \& Dickinson 1998;
van Dokkum et al. 1998b; Barger et al. 1998; Broadhurst \& Bouwens 1999),
but are brighter overall (van Dokkum \& Franx 1996), suggesting an early
formation epoch $z \gta 2$ for an $\Omega_m = 0.3$, 
$\Omega_{\Lambda} = 0.7$ universe (van Dokkum et al. 1998a).

Set against this conventional picture are a number of 
recent studies that suggest that most or 
many ellipticals have formed continuously over
time by mergers that introduce morphological disturbances 
(Schweizer \& Seitzer 1992) and new generations of 
young stars (Kauffmann \& Charlot 1998).
Some observations of stellar photometric indices 
(Prugniel, Golev, \& Maubon 1999) 
and deficiencies of E galaxies at redshifts $z \lta 1$ 
(Kauffmann, Charlot \& White 1996) 
support significant ongoing evolution, 
particularly for field ellipticals,
but other studies at these redshifts 
indicate a nearly constant E/S0 space density 
(Im et al. 1996: Franceschini et al. 1998) and old formation ages
(Bernardi et al. 1998).

The age of stars in ellipticals has also been questioned, but progress
has been confounded by the age-metallicity conspiracy:
youthful, metal-rich and old, metal-poor populations are nearly 
indistinguishable (Worthey 1997).
However, Lick observers have demonstrated that Balmer 
absorption lines can partially break this 
age-metallicity degeneracy (Gonzalez 1993; Faber et al. 1995; Trager 1997).
By comparing the H$\beta$ equivalent width (index) and a Mg + Fe photometric
index with expectations from Worthey's evolutionary models, Gonzalez found
a wide range of apparent ages in a sample of 40 (mostly field) ellipticals
for which the mean age is only 
$8 \pm 3$ Gyrs.
In addition, stars in the inner regions of these galaxies 
($r \lta r_e/8$) are about 3 Gyrs younger (and
more metal-rich) than those at larger radii.
These stellar ages are inconsistent with 13-15 Gyrs, formerly
thought to be more likely. The age spread is larger for ellipticals
of lower luminosity. Most of the H$\beta$ equivalent width is contributed 
by F and G dwarf stars near main sequence turnoff.

\section{CONTINUOUS STAR FORMATION}

In a series of recent papers we have presented detailed models of the 
evolution of elliptical galaxies with
an emphasis on the gas dynamics of the hot interstellar gas 
(e.g. Brighenti \& Mathews 1999a; 1999b). Galactic stars are assumed
to form at $t_{*s} = 1$ Gyr and the de Vaucouleurs structure of the large
ellipticals is constructed at $t_* = 2$ Gyrs. Our calculations successfully 
reproduce currently observed interstellar (cooling flow) density,
temperature and (iron) abundance profiles in massive ellipticals with a minimum
of adjustable parameters.

The global rate that interstellar gas cools 
can be estimated by dividing the observed 
(bolometric) X-ray luminosity $L_x$ by the enthalpy per gram of the 
gas, ${\dot M} = (2 \mu m_p / 5 k T)L_x \approx 2.5$ $M_{\odot}$ 
yr$^{-1}$ where $T = 1.3 \times 10^7$ K is typical of large ellipticals 
($m_p =$ proton mass; $\mu = 0.62 =$ molecular weight). The total mass of gas
that cools over cosmic time, several $10^{10}$ $M_{\odot}$, 
is only about 4 - 5 percent of the total baryonic mass currently
in stars.

Two of the most perplexing and long standing problems concerning 
galactic cooling flows are (1) to determine where cooling to low temperatures 
actually occurs in the galaxies and (2) 
to determine the final physical disposition of the cooled gas.
The dropout or cooling of interstellar gas must occur over a substantial
volume of the inner galaxy, but the radial mass profile of 
cooled gas cannot be predicted from first
principles since it depends critically on entropy fluctuations acquired
during a variety of complex processes (stellar mass loss,
supernovae explosions, magnetic field variations, etc.).
To accommodate this uncertainty, we have considered a variety of cooling 
dropout models in which the hot gas is 
assumed to cool at a rate 
$(\partial \rho / \partial t)_{do} = -q(r) \rho/t_{do}$, 
where $t_{do} = 5 m_p k T / 2 \mu \rho \Lambda$ is the local
(constant pressure) cooling time and $q(r)$ is an adjustable dropout 
function (Brighenti \& Mathews 1999b).
We compare computed interstellar properties with currently
($t_n = 13$ Gyrs) observed interstellar gas in the luminous elliptical 
NGC 4472. 
Most of these models are unacceptable because the radial distributions of 
X-ray surface brightness $\Sigma_x(r)$, gas density $n(r)$ or 
temperature $T(r)$ disagree with profiles observed in this galaxy. 
Among the models considered, the simple constant $q(r) = 1$ model
gave the best results, although the agreement with 
observed $\Sigma_x(r)$ was still not perfect;
we consider this model again here to estimate the mass dropout
in NGC 4472.

Regarding the second perplexing and long standing problem, 
it has long been speculated that the end
product of the cooled gas are low mass,
non-luminous stars (Fabian, Nulsen \& Canizares 1982;
Thomas 1986; Cowie \& Binney 1988; 
Vedder, Trester \& Canizares 1988;
Sarazin \& Ashe 1989; Ferland, Fabian \& Johnstone 1994).
We have recently reconsidered the star formation process
in elliptical galaxy cooling flows and have concluded that the mass of stars
in the dropout stellar population probably extends to
$\sim 2$ $M_{\odot}$, i.e. {\it the dropout population 
is optically luminous} (Mathews \& Brighenti 1999a).
As cold gas collects at a cooling site, it becomes 
gravitationally unstable at this limiting mass, setting a firm upper mass
limit $m_u$ on the IMF for stars forming in the 
central regions of massive ellipticals.
The upper mass limit on the (bottom-heavy) IMF increases
only modestly with galactic radius 
($\sim 4$ $M_{\odot}$ at $r \approx r_e$) and is almost independent 
of time during the evolution of the cooling flow
for redshifts $z \lta 1$.

Additional support for the formation of optically 
luminous stars in galactic cooling 
flows is provided by the thinness of the fundamental plane.
For agreement with observed X-ray surface brightness distributions 
in ellipticals, most of the mass of cooled gas is  
concentrated well within $r_e$
where the relatively small
dropout mass can contribute substantially to 
the central mass and mass to light ratio determined from
stellar velocities.
If the dropout stellar population is assumed 
to be non-luminous, Mathews \& Brighenti (1999b) 
have shown that the variation of dark 
dropout mass among ellipticals causes large,
undesirable shifts in the fundamental plane that are incompatible
with its observed thinness.
However, these perturbations on the fundamental plane may be
lessened or removed if the dropout stars are luminous.

\section{$L_B$, $B-V$ AND $H\beta$}

Star formation in ellipticals is efficient in the sense that the total 
mass of HII gas and cold neutral or molecular gas at 
any time are both much less than the total mass of gas that has cooled.
Therefore, the star formation rate ($\Psi_{SFR}$) in NGC 4472 
is equal to the instantaneous 
rate that hot interstellar gas cools by radiative losses.
The total accumulated mass that has cooled in NGC 4472 since 
$t_* = 2$ Gyr, $M_{do}(t)$, and $\Psi_{SFR} = dM_{do}/dt$ are shown 
in Figure 1a; these are based on the $q = 1$ model 
that best fits the X-ray observations 
of NGC 4472 (Brighenti \& Mathews 1999b).
The mass dropout $M_{do}(r,t_n)$ at $t_n = 13$ Gyrs for this 
model occurs mostly in $r \lta r_e$ and the total dropout mass 
is $M_{do}(t_n) = 4.7 \times 10^{10}$ $M_{\odot}$.
This is much less than the total current stellar mass in NGC 4472,
$M_{*t} = 7.26 \times 10^{11}$ $M_{\odot}$, determined with 
$M/L_B = 9.2$.

For the purpose of illustration, we assume that the old galactic 
stars can be approximated
as a single burst stellar population having a Salpeter IMF 
from $m_{\ell} = 0.1$ to $m_u = 125$ $M_{\odot}$. 
By contrast, the younger dropout stellar population with variable SFR,
$\Psi_{SFR}(t)$, is assumed to have a Salpeter IMF from 
$m_{\ell} = 0.1$ to $m_u = 2.5$ $M_{\odot}$.
Both of these IMFs are available in single burst
format in the 1999 Bruzual-Charlot library for isochrone
synthesis spectral evolution 
(Charlot \& Bruzual 1991: Bruzual \& Charlot 1993), 
assuming solar abundance. The total B-band luminosity of the
dropout stellar population is
$$ L_{B,do}(t) = \int_{0}^{t - t_*} \Psi_{SFR}(t-\tau) 
\ell_{B,do}(\tau) d \tau$$
where $\ell_{B,do}(t)$ is the single burst B-band luminosity 
per unit solar mass.
The luminosity of the background old single 
burst population is 
$L_{B,old}(t) = f_m M_{*t} \ell_{B,old}(t)$ 
where $f_m$ is 
a coefficient of order unity that must be adjusted for agreement
with the observed B-band luminosity of NGC 4472 (see below).

The evolution of B-band luminosities and B-V colors are illustrated 
in Figure 1b and 1c for each stellar population and for their 
combined radiation. For reference, the redshift $z = 1$ is shown 
at time $t = 6.19$ Gyrs (assuming $H_o = 65$ km s$^{-1}$, 
$\Omega_m = 0.3$ and $\Omega_{\Lambda} = 0.7$).
We have chosen $f_m = 1.35$ so that the total B-band luminosity 
$L_{B,tot} = L_{B,old} + L_{B,do} = 7.89 \times 10^{10}$ 
$L_{B,\odot}$, appropriate for NGC 4472 at distance 
$d = 17$ Mpc.
Although the dropout population currently contributes about
15 percent of the total B-band light 
($L_{B,do}(t_n) = 1.2 \times 10^{10}$ $L_{B,\odot}$),
its fractional contribution to 
the galactic light is quite constant for redshifts $z \lta 1$.
The combined do+old population is only slightly bluer 
(by $\delta(B-V) \sim 0.03$) than the old population and this 
difference is essentially constant for $z \lta 1$.
Formally, the $B - V$ of the combined population indicates
an age $\sim 8.5$ Gyrs that is less than that of the old
population alone, 12 Gyrs, but metallicity variations 
could produce a similar color variation.

To estimate the equivalent width of H$\beta$ we assume 
that the line width is similar for both populations:
$$EW_{\beta,tot}(t) = {
\langle ew_{\beta} \ell_{B} \rangle_{do}(t) 
+ f_g ew_{\beta,old}(t)L_{B,old}(t)  \over
L_{B,do}(t) + f_g L_{B,old}(t) }$$
where
$$\langle ew_{\beta} \ell_{B} \rangle_{do}(t) =
~~~~~~~~~~~~~~~~~~~~~~~~~~~~~~~~~~~$$
$$\int_0^{t - t_*} \Psi_{SFR}(t - \tau) 
\ell_{B,do}(\tau) ew_{\beta,do}(\tau) d \tau.$$
Here $f_g(R)$ is the ratio of light from the old 
to dropout population within projected radius $R$ 
normalized to the total ratio of old to dropout light; 
generally we assume $f_g = 1$, corresponding to 
viewing the total light from both populations.

As an illustration, we use single burst 
$ew_{\beta}(t)$ from the 1999 Bruzual-Charlot (BC99) 
tables appropriate to the IMF of each 
population. As shown in Figure 2 the dropout population 
(with $f_g = 1$) reduces the apparent age 
of the old population by
$\sim 5$ Gyrs, 
i.e. $EW_{\beta,tot}(t_n) 
\approx EW_{\beta,old}(t_n - 5~{\rm Gyrs})$,
in agreement with observations that 
correlate bluer colors with stronger 
H$\beta$ (Forbes \& Ponman 1999).
For smaller $f_g = 1/4$, corresponding to viewing 
NGC 4472 within $r_e$, the apparent age is reduced 
by $\sim 8.5$ Gyrs.
Actual observations of galactic cores 
view a fraction of both populations, i.e. the apparent 
H$\beta$ age is aperture-dependent.
For $m_u \gta 2$, $ew_{\beta,do}(t)$  
and $EW_{\beta,tot}(t_n)$ are insensitive 
to $m_u$, i.e. $ew_{\beta,do}(t) = 
ew_{\beta,old}(t)$ can be assumed. 
We also used the population code 
available at the Worthey website to
determine both $EW_{\beta,tot}(t_n)$ and 
$(B - V)_{tot}(t_n)$ for a Salpeter dropout
IMF from 
$m_{\ell} = 0.2$ to $m_u = 10$ $M_{\odot}$.
For this model the H$\beta$ and $(B - V)$ 
ages are 8.5 and 9.5 Gyrs with $f_g = 1$. 
These age uncertainties are consistent with 
($\sim 35$ \%) errors inherent to population synthesis
procedures
(Charlot, Worthey \& Bressan 1996; Worthey 1996).

Clearly, however, the contribution of cooling dropout
stars to the spectra of elliptical galaxies 
can explain the relatively young ages inferred from 
H$\beta$ observed in some ellipticals 
even if the underlying stars are very old.

\section{FURTHER DISCUSSION}

All massive ellipticals contain cooling interstellar 
gas so no comparisons with gas-free galaxies can be made.
However, since the cooling flow mass dropout is centrally
concentrated within $r_e$, the equivalent width of 
H$\beta$, $EW_{\beta}$, should increase toward galactic centers,
in agreement with the observations of Gonzalez (1993).
Dropout star formation should be accompanied by 
an ensemble of additional observations at small galactic radii: 
H$\beta$ {\it in emission} from cooling clouds, enhanced X-ray 
surface brightness due to dense, locally cooling
regions, and lower apparent X-ray temperatures 
which (with hydrostatic equilibrium) indicate
interior masses {\it less} than the known stellar mass.
Such mass discrepancies within $\sim 0.1r_e$ are 
apparent in X-ray observations of bright Virgo ellipticals
(Brighenti \& Mathews 1997a). Some fraction of the total 
optical light ($\sim 15$ percent in Figure 1) 
in E galaxies comes from the dropout population
with $(M/L_B)_{do} \approx 4 < (M/L_B)_{old}$.
Therefore, $(M/L_B)$ should vary with galactic radius and
produce a shift away from the fundamental plane 
defined by the old stars 
alone.

The observed apparent H$\beta$ age spread among ellipticals is large:
2 to 12 Gyrs in the Gonzalez (1993) sample, 5 to 12 Gyrs in the 
Fornax cluster (Kuntschner \& Davies 1998) and 8 to 12 Gyrs 
in the Coma cluster (Jorgensen 1999). Such variations
can be expected if the cooling dropout profile $q(r)$ 
differs among otherwise similar galaxies, 
represented here with the factor $f_g$.
But some of the age variation may arise from
comparing ellipticals of greatly different $L_B$.  Interstellar X-ray
emission from ellipticals with $L_B \lta 3 \times 10^{10}$ is masked
by stellar X-rays, but cooling flows still exist in these faster
rotating, low $L_B$ ellipticals.  We have shown (Brighenti \& Mathews
1997b) that large cold gas disks may form from cooling flow gas in low
$L_B$ ellipticals similar to the HI disks observed by Oosterloo,
Morganti \& Sadler (1999).  Occasional star formation with normal IMFs
and maximal $EW_{\beta}$ (de Jong \& Davies 1997) is therefore
expected in low $L_B$ ellipticals.  In luminous ellipticals ($L_B \gta
3 \times 10^{10}$) $m_u$ and $EW_{\beta}$ will generally be lower due
to isolated star formation in the high pressure ISM. But $L_x/L_B$
varies enormously among bright ellipticals of similar $L_B$ and
H$\beta$
ages should reflect this same variation since the fraction of mass in
the dropout population is proportional to $L_x/L_B$.  As expected,
ellipticals in Gonzalez' sample with $L_B > 3 \times 10^{10}$ and low
$L_x/L_B$ -- NGC 4649, NGC 7619 and NGC 7626 -- also appear to be very
old ($\sim 13$ Gyrs).  However, this interpretation is not entirely
straightforward; ellipticals with large $L_x/L_B$ also have larger
interstellar pressure which may result in $m_u < 2$ and lower H$\beta$
indices.  Although H$\beta$ ages are insensitive to the IMF of the
dropout population for $m_u \gta 2.5$ $M_{\odot}$, as $m_u$ approaches
$\sim 0.8 M_{\odot}$, $EW_{\beta}$ decreases; this is just the range
in $m_u$ anticipated from our model of star formation in luminous
ellipticals (Mathews \& Brighenti 1999a).  The H$\beta$ index may not
vary monotonically with $L_x/L_B$.

Although the star formation process described here obviates the need for
continued galactic merging to account for the H$\beta$ equivalent widths 
observed, we do not claim that recent merging in ellipticals is non-existent 
or unimportant. We do note, however, that rather few images of ellipticals 
at small redshift indicate ongoing mergers with 
gas-rich, star-forming galaxies. But if such
regular mergers do generate young stellar populations in ellipticals, 
it may be possible to detect azimuthal asymmetries in the stellar 
H$\beta$ that reflect the orbital plane(s) of the 
newly-introduced stars. Young stars formed from cooling flow dropout
are expected to be symmetrically disposed in the galactic
potential, but their orbits may be more radial with narrower lines than those
of the old stellar population.

\acknowledgments
We are indebted to Stephane Charlot for providing H$\beta$ indices 
for the two populations considered and to 
Stuart Norton for helpful advice.
Studies of the evolution of hot gas in elliptical galaxies 
at UC Santa Cruz are supported by
NASA grant NAG 5-3060 and NSF grant 
AST-9802994 for which we are very grateful. 
FB is supported in part by Grant MURST-Cofin 98.








\vskip.1in
\figcaption[aasbhfig1.ps]{
(a) {\it Solid line:} Cumulative mass of cooled 
(dropped-out) gas in NGC 4472 in units of 
$10^{10}$ $M_{\odot}$;
{\it Dashed line:} Star formation rate 
($M_{\odot}$ yr$^{-1}$)
from cooled interstellar gas in NGC 4472.
(b) B-band luminosity in $10^{10}$ $L_{B,\odot}$ 
for the dropout stellar population ({\it short dashed line}), 
the old stellar population ({\it long dashed line}) 
and both populations combined ({\it solid line}).
(c) $(B - V)$ colors of the old stellar population 
({\it long dashed line}) and both populations combined 
({\it solid line}).
The {\it dotted line} is drawn parallel to the horizontal axis.
Redshifts $z = 0$ and 1 are indicated for a cosmology 
having $H_o = 65$ km s$^{-1}$, $\Omega_m = 0.3$ and 
$\Omega_{\Lambda} = 0.7$.
\label{fig1}}

\vskip.1in
\figcaption[aasbhfig2.ps]{
H$\beta$ equivalent widths in Angstroms for the dropout 
stellar population ({\it short dashed line}),
the old stellar population ({\it long dashed line}) 
and both populations combined ({\it solid line}).
The {\it dotted line} is drawn parallel to the horizontal axis.
Redshifts are labeled as in Figure 1.
\label{fig2}}

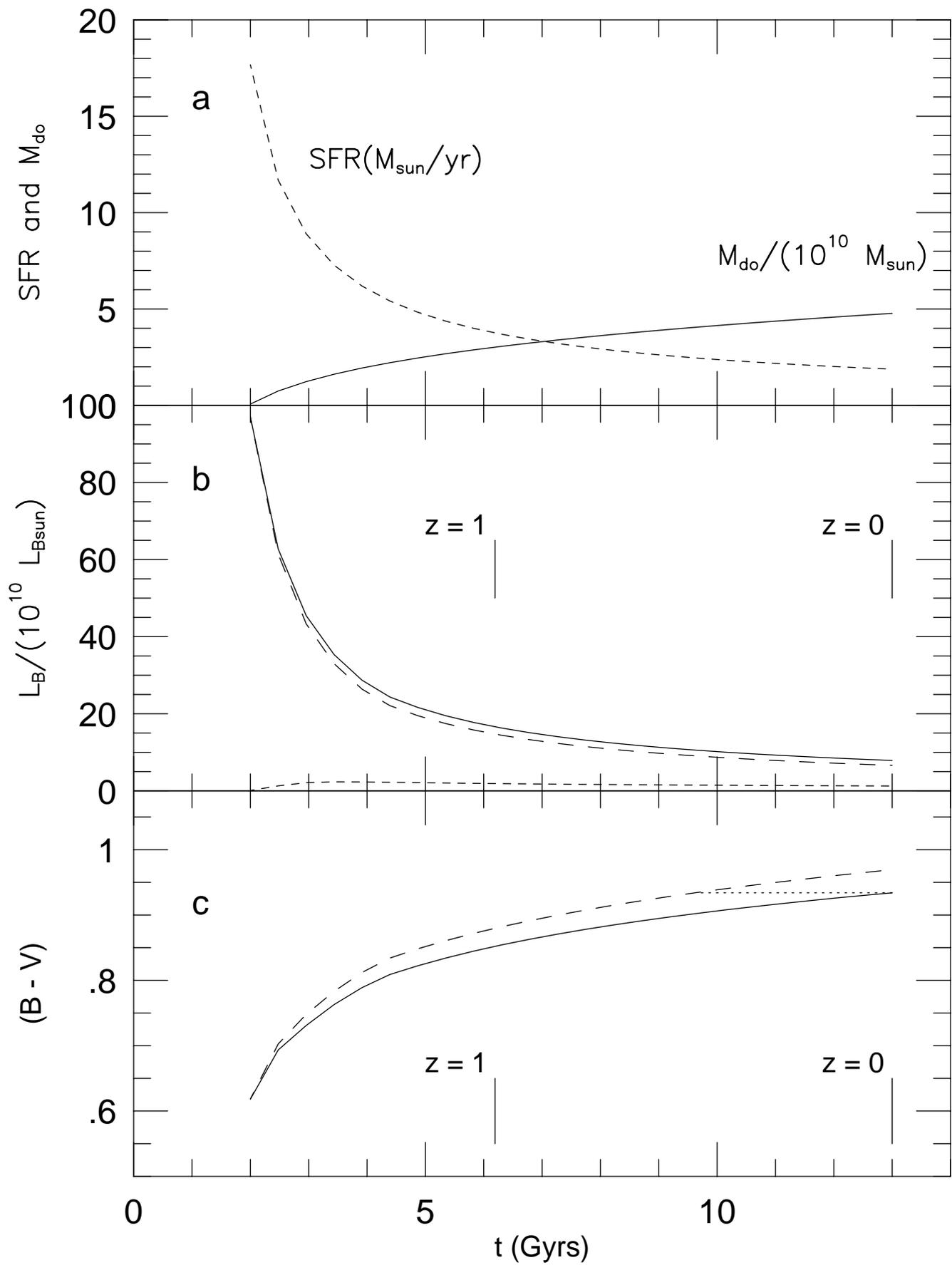
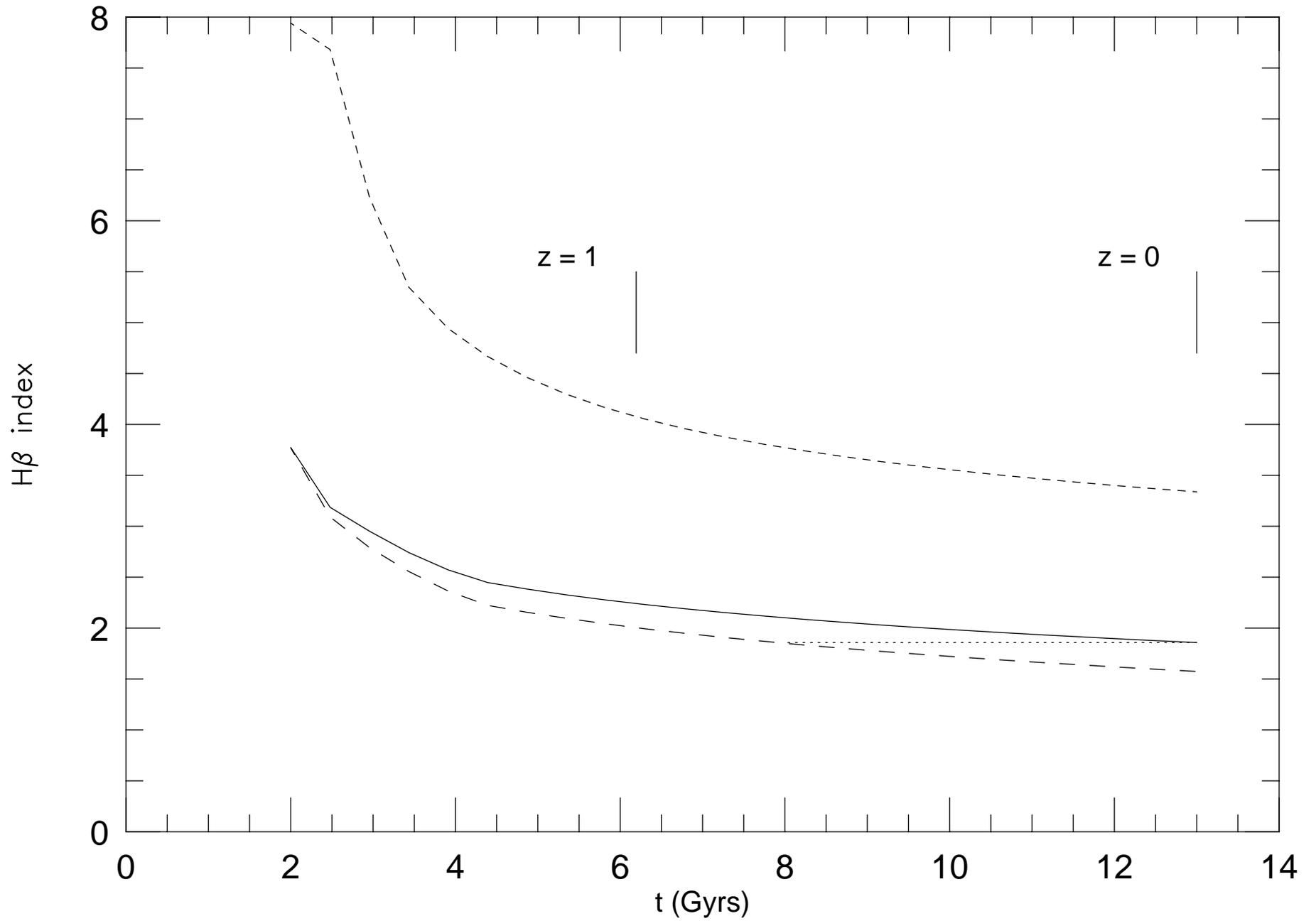
\end{document}